\documentclass[orivec,a4paper]{llncs}

\pdfoutput=1

\usepackage[latin1]{inputenc}
\usepackage{amsmath}
\usepackage{amssymb}
\usepackage{epsfig}
\usepackage{url}
\usepackage{xcolor}
\usepackage{graphics,graphicx}
\usepackage{tikz}
\usetikzlibrary{positioning,fit,shapes.geometric,shapes.arrows,calc}
\usepackage{hyperref}
\usepackage[paper=a4paper,left=4cm,right=4cm]{geometry}

\author{Barbara Kordy, Piotr Kordy, Sjouke Mauw, Patrick Schweitzer\\
\texttt{\{barbara.kordy, piotr.kordy, sjouke.mauw, patrick.schweitzer\}@uni.lu}}
\title{ADTool: Security Analysis with Attack--Defense Trees}
\subtitle{Extended Version\footnote{This is an extended version of the 
tool demonstration paper accepted for publication at QEST 2013~\cite{qest}.
The research leading to 
the results presented in this work received 
funding from the Fonds National de la 
Recherche Luxembourg under the grants C08/IS/26 and PHD-09-167 and 
the European Commission's Seventh Framework Programme
(FP7/2007-2013) under grant agreement number~318003 (TREsPASS).
}}
\institute{University of Luxembourg, SnT\\
6, rue Richard Coudenhove-Kalergi, 
L-1359 Luxembourg
}

\begin{document}

\maketitle

\begin{abstract}
The ADTool is free, open source software assisting graphical modeling and 
quantitative analysis of security, using attack--defense trees. 
The main features of the ADTool are easy creation, efficient 
editing, and automated bottom-up evaluation of security-relevant measures. 
The tool also supports the usage of attack trees, protection trees and defense 
trees, which are all particular instances of attack--defense trees.
\end{abstract}

\section{Introduction} 
\label{sec:introduction}

Graphical security models provide an intuitive but systematic methodology to 
visualize possible attacks and countermeasures and to enable the computation 
of various security related parameters. Thus, they have been widely accepted 
by the industrial sector, as a means to support and facilitate threat 
analysis and risk management processes. Graphical models have been used for 
security analysis of SCADA systems \cite{ByFrMi,TeLiGo}, of voting 
systems~\cite{LaDiEpHa,BuTr}, of vehicular communication 
systems~\cite{HeApFuRoRuWe,AiBoDoFeGeKrLe}, of Internet related
attacks~\cite{TiLaFiHa,LiZaRuLi}, and of socio-technical
attacks~\cite{BaKoMeSc12,EoPaPaCh,ReVeOlCu}. 
This paper presents the ADTool 
software~\cite{ADToolPP} which supports 
quantitative and qualitative security assessment using 
a graphical security modeling technique called 
attack--defense trees. 

\subsection{Background}

Attack--defense trees (ADTrees)~\cite{KoMaRaSc}
extend and improve the well-known formalism of attack trees, by 
including not only the actions of an attacker, but also 
possible counteractions of a defender. 
Hence, an ADTree can be seen as a game between two opposite
players: an attacker trying to compromise a system and a defender trying to
protect the system~\cite{KoMaMeSc}. 
The root of the tree represents the main 
goal of one of 
the players. Using an AND-OR tree structure, this main goal is then refined 
into less complex subgoals that need to be reached in order to achieve 
the objective of the root. Every node of an ADTree can 
also be counteracted. This is depicted using a subtree rooted in a node of the 
opposite player. 
Since interactions between an attacker and a 
defender are modeled explicitly in ADTrees, 
the extended formalism allows for a more 
thorough and accurate security analysis compared to attack trees, 
without however requiring additional computational power~\cite{KoPoSc}. 
A toy example of an ADTree, representing how to attack a 
server, is given in Figure~\ref{fig:Attack_on_server}. 

Theoretical foundations of the ADTree methodology, including 
a graphical and a term-based syntax as well as numerous 
formal semantics, have been introduced in~\cite{KoMaRaSc}.
A mathematical framework for quantitative evaluation of 
ADTrees has been developed in~\cite{KoMaRaSc2}.
It is based on the notion of attributes, which allow us to formalize
and specify relevant security metrics. 
Standard quantitative analysis of ADTrees relies on a step-wise computation 
procedure. Numerical values are assigned to all atomic actions, 
represented 
by the non-refined nodes\footnote{Unlike in attack trees, 
a non-refined node of an ADTree is not necessarily a leaf. 
Such a node does not have any refining children but can still be countered.}.
The values for the 
remaining nodes, including 
the root of the tree, are deduced automatically in a bottom-up way. 
This bottom-up algorithm makes use of 
attribute domains 
which specify operators 
to be used while calculating values for 
different node configurations of an ADTree. 
In \cite{KoMaSc}, we have developed guidelines for 
intuitive and formal specification and classification
of the most popular measures for attack trees and ADTrees. 
There, we also provide
guidelines for how to properly specify an attribute, so that it can be 
evaluated on ADTrees using the bottom-up algorithm. 

\subsection{Motivation}

While validating the ADTree approach on a large industrial case 
study~\cite{BaKoMeSc12}, it became apparent that the practical use of the 
methodology requires dedicated tool support. Lack of such support 
may result in numerous modeling difficulties and computational errors. 
On the one hand, there exist a number of commercial software applications for 
attack tree-like modeling, including 
SecurITree~\cite{Program1}, 
and AttackTree+~\cite{Program2}. 
However, these are closed source tools and their use is not free of charge. 
On the other hand, existing academic software, such as 
SeaMonster~\cite{Program6}, 
does not support quantitative analysis and uniformly integrated defenses.

The above observations motivated the development of the ADTool, which
\begin{itemize}
\item is a free and open source application;
\item supports quantitative and qualitative analysis of tree-based models 
integrating attack and defense components;
\item is based on well-founded formal framework;
\item guides the user in constructing syntactically correct 
models; 
\item allows to draw visually appealing trees;
\item facilitates sharing, management and updating of the models;
\item helps in verification of numerical values provided by the users;
\item automates computation of security related parameters.
\end{itemize}

This paper provides an overview of the main 
features, practical capabilities, and the architecture of the ADTool.  
For a complete description of  how to use the tool, we refer 
to the ADTool  manual~\cite{ADTool_Manual}.

\section{Main features of the ADTool}
The goal of the ADTool is to  provide security consultants 
as well as academic researchers with a rigorous but user-friendly application 
that supports security analysis and risk assessment process 
using attack--defense trees. 
It integrates two crucial modeling aspects:
the creation of security models and their quantitative analysis.
From a formal perspective, 
attack trees~\cite{Schn,MaOo}, protection trees~\cite{EdDaRaMi},
and defense trees~\cite{BiFiPe} are instances of attack--defense trees.
Thus, the ADTool can also be employed to automate and facilitate the usage of
all aforementioned formalisms.  

\subsection{Security modeling using ADTool}

\begin{figure}[bhtp]
\centering
\includegraphics[width=1\textwidth]{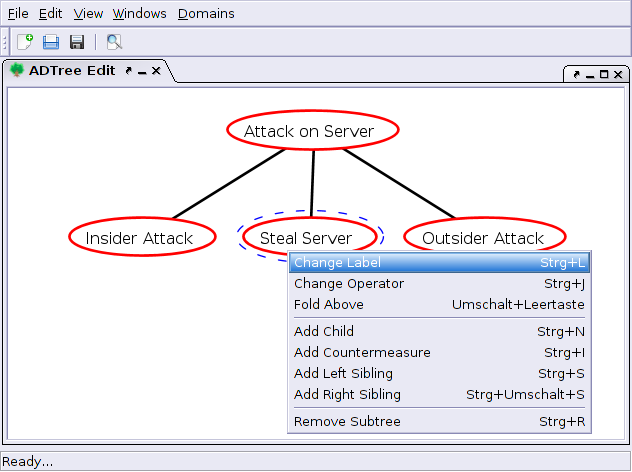} 
\caption{Creating an ADTree in ADTool.}
\label{fig:Menu}
\end{figure}

One of the main features of the ADTool is 
its user-friendliness. 
The tool is guiding the user in constructing
models that comply with the
graphical ADTree language, as described in \cite{KoMaRaSc2}. 
All options that allow to modify or
refine a given component of a model can be accessed by right-clicking 
the node, as shown in Figure~\ref{fig:Menu}. 
Alternatively, intuitive keyboard shortcuts 
can be used to create, alter or remove a subtree~\cite{ADTool_Manual}.

\begin{figure}[htb]
\centering
\includegraphics[width=\textwidth]{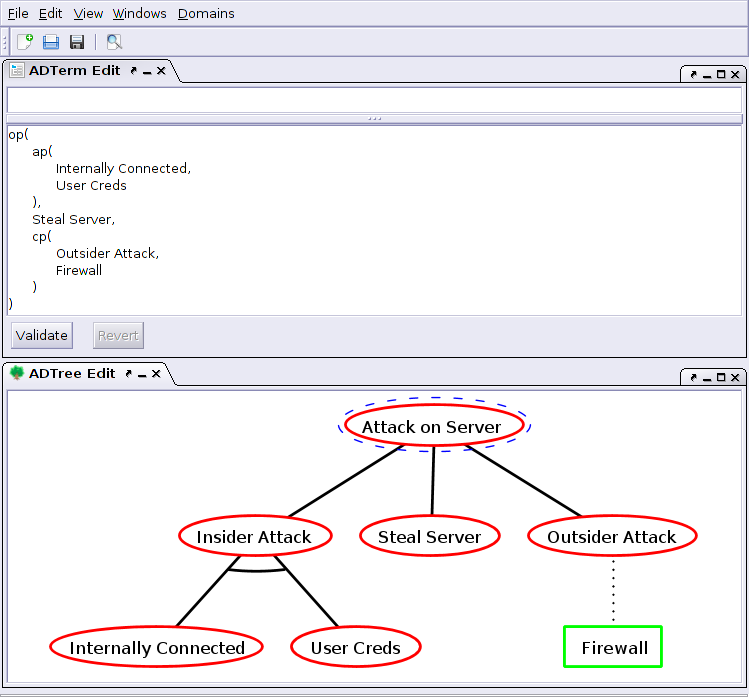} 
\caption{An ADTree modeled in ADTool.}
\label{fig:Attack_on_server}
\end{figure}

An improved version of Walker's algorithm~\cite{Walk}, with enhancements 
suggested by Buchheim~et~al.~\cite{BuJuLe},
has been implemented in the ADTool to produce 
trees having an appealing layout.
Furthermore, when an ADTree is built, the corresponding 
attack--defense term (ADTerm) is immediately
displayed, 
as shown in Figure~\ref{fig:Attack_on_server}. 
ADTerms form a compact, algebraic representation of ADTrees.
In order to link a term  model with its graphical counterpart, 
the shortest tree edit distance algorithm~\cite{DeMoRoWe,PaAu}
has been implemented. 
It ensures that when an ADTerm is modified, the corresponding 
ADTree is adapted accordingly. 

\begin{figure}[htb]
\centering
\includegraphics[width=\textwidth]{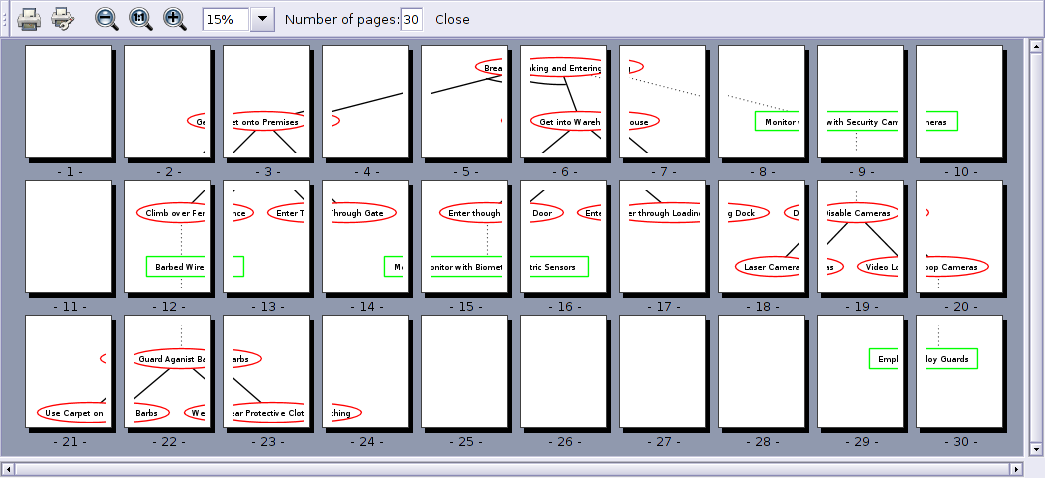} 
\caption{Large-scale printing using ADTool.}
\label{fig:printing}
\end{figure}

Finally, the ADTool provides advanced features for model manipulation and 
management.  
Folding, expanding and zooming options make the analysis of large models 
possible. Temporarily hiding parts of a tree
permits users to focus on the displayed components. 
This is highly appreciated during industrial meetings and presentations. 
ADTrees created with the ADTool can be saved as special~.adt files, 
which enables their reuse and modification.
Models can also be exported to vector graphics files (pdf), 
raster graphics files (png, jpeg) and \LaTeX\ files (tex). 
Resulting figures can be used as illustrations in 
scientific and industrial 
presentations, research papers and posters. 
A dedicated option, illustrated in Figure~\ref{fig:printing}, 
makes it possible 
to print trees on a specified number of pages, 
which enhances readability of large-scale models. 

\subsection{Quantitative analysis using ADTool}

\begin{figure}[htbp]
\centering
\includegraphics[width=\textwidth]{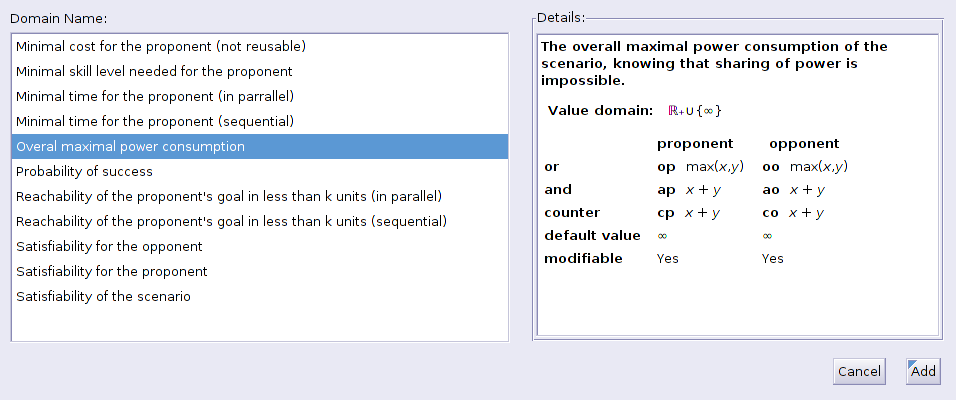} 
\caption{Attribute selection in ADTool.}
\label{fig:attribute_domain_selection}
\end{figure}

The bottom-up algorithm for evaluation of 
attributes on ADTrees has been implemented in the ADTool. 
Supported measures include: 
attributes based on real values (e.g., time, cost, probability), 
attributes based on levels (e.g., required skill level, reachability 
of the goal in less than $k$ units of 
time),
and Boolean properties (e.g., satisfiability of a scenario).
The implemented measures can be computed from the point of view of 
an attacker (e.g., the cost of an attack), of a defender (e.g., 
the cost of defending a system), or relate to both of them (e.g., overall 
maximum power consumption). 
By using different attribute domains, 
we distinguish between situations when actions are executed simultaneously and 
when they  need to be performed sequentially.  
The choice of an attribute is illustrated in Figure~\ref{fig:attribute_domain_selection}.

\begin{figure}[htb]
\centering
\includegraphics[width=\textwidth]{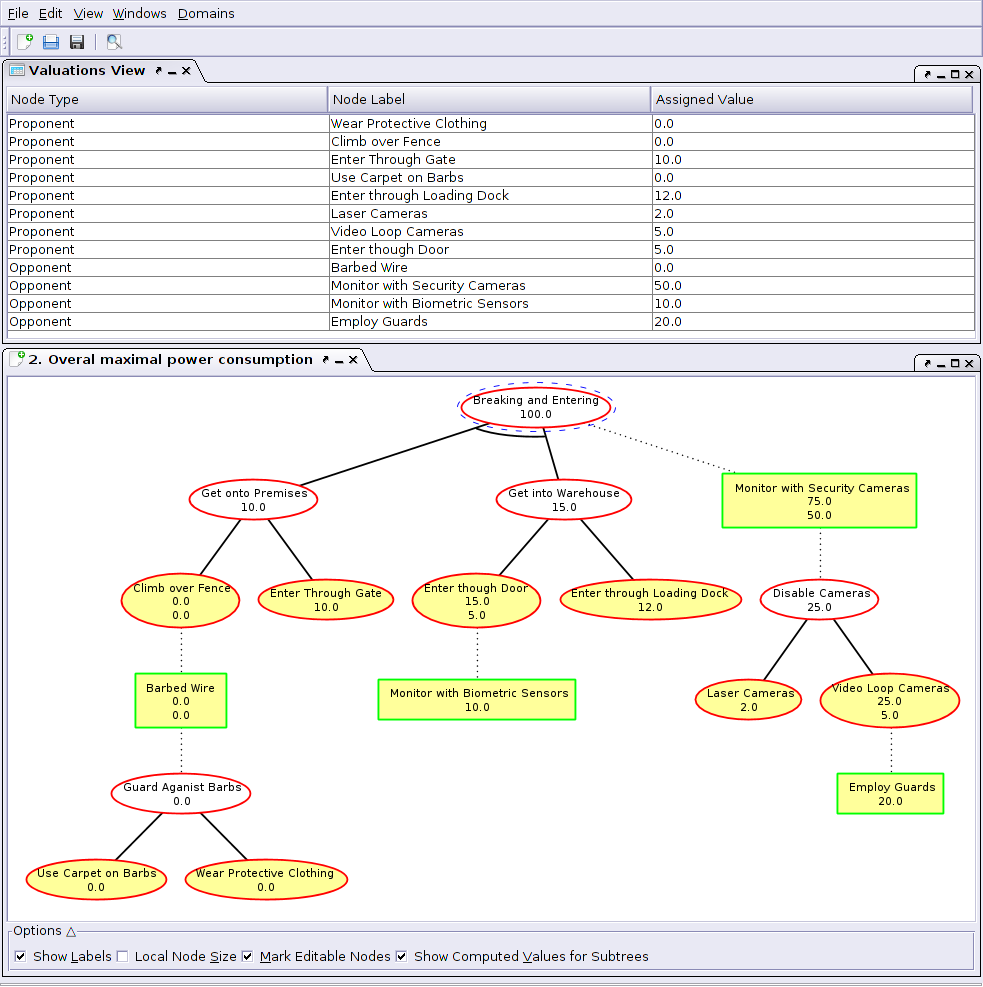} 
\caption{Attribute evaluation in the ADTool.}
\label{fig:attribute_values}
\end{figure}

After a user has selected an attribute, 
the tool decorates the ADTree with default values representing the 
worst case scenario, e.g., infinite cost or maximal required skill level.
The user then  
customizes the inputs for the relevant non-refined nodes directly on the tree
or uses an overview table, as shown in Figure~\ref{fig:attribute_values}.
The use of the table is particularly helpful in case of large models. 
The tool ensures that the provided values are consistent and belong to 
a specified value domain.
This is especially important when several specialists supply values for
different parts of the tree.  Nodes labeled with the same name, 
i.e., representing the same action, automatically receive the same value. 
Such a design choice is consistent with the ADTree 
methodology, as specified in~\cite{KoMaRaSc2}.
If a value for a non-refined node is modified, 
ADTool automatically computes the values of the remaining nodes using 
the bottom-up algorithm. 
By restricting the user input to the minimum and by automating 
computations, we avoid calculation errors. 

\subsection{Evaluation of ADTool}

The ADTool easily handles trees containing a few thousand nodes.
The complexity of the bottom-up algorithm for evaluation of attributes 
on ADTrees is linear with respect to the size of the tree, i.e., 
its number of nodes, and the computations are performed instantaneously. 
Thus, the quantitative analysis using ADTool scales very well. 
The limiting factor of the application is the graphical display of the models.
For trees of more than ten thousand nodes, a delay of about 
five seconds occurs when a new node is added.
This is due to the recalculation of the positions of some nodes. 
ADTool is programmed in  Java. Implementation in C++
would run faster, but its development overhead would be much higher.

ADTool presents a number of advantages compared to existing tools employed for 
attack tree-based modeling. Graphical security models 
are usually prepared with the help of non-dedicated drawing tools. 
In this case, the user himself needs to make sure that the models he 
creates are syntactically correct. 
Since ADTool guides the user in creating only well-formed models, 
its employment does not require an in-depth knowledge of the technical 
details of the underlying methodology. 

Another strong point of our application is that it supports both:
creation and management of graphical models as well as their quantitative 
analysis. 
Existing prototype tools usually concentrate on one of the above aspects only.
On the one hand, applications such as 
SeaMonster~\cite{Program6} support creation of 
security models, but they do not
provide means to perform computations. On the other hand, most academic 
tools designed for quantitative analysis are command-line programs and 
they do not support visual representation of the analyzed models. 

Finally, the ADTree methodology and the computation procedures implemented in 
ADTool have sound theoretical foundations~\cite{Website_ATREES}. 
In particular, the meaning of all attributes supported by the tool is clearly 
defined and the underlying mathematical models are well-studied from the 
formal point of view~\cite{KoMaRaSc,KoMaRaSc2,KoMaSc}. 
Together with the fact that ADTool is a free and open source application, 
this makes it a good alternative for commercial software
for attack tree-based modeling, such as~\cite{Program1,Program2}, etc.

\section{Implementation Characteristics} 

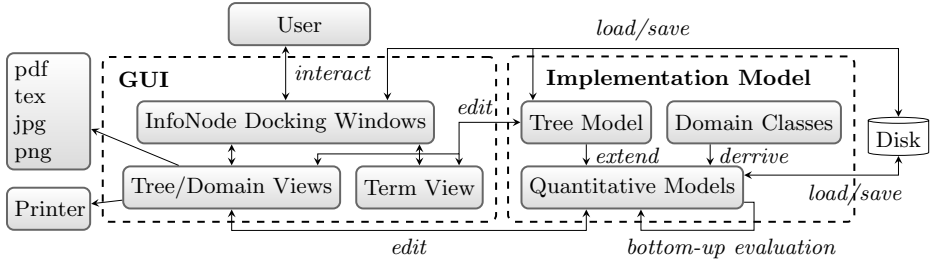
\begin{figure}[bhtp]
\scalebox{0.93}{\begin{tikzpicture}[
   actor/.style={
   rectangle, minimum size=6mm, rounded corners=1mm, thick,
   draw=black!60, top color=white,bottom color=black!14},
   collection/.style={rectangle, minimum size=1cm, thick, dashed, rounded
   corners=1mm, draw},
   >=stealth,
   vertical con/.style={stealth-stealth,
   to path={
     (\tikztostart|-\tikztotarget.south) -> (\tikztostart.north)\tikztonodes
   }},
   ]
   \node(info) [actor] {InfoNode Docking Windows};
   \node(tree) [actor,anchor=north west,yshift=-0.3cm,xshift=-0.2cm] at (info.south west)
   {Tree/Domain Views};
   \node(term) [actor,right=0.2cm of tree] {Term View};
   \draw[vertical con](term) to (info);
   \draw[vertical con](tree) to (info);
   \node(gui) [above,anchor=south west,xshift=-0.3cm,inner sep=1.0pt,yshift=4pt] at (info.north west){\textbf{GUI}};
   \node(guiall) [collection,fit=(tree)(term)(gui)(info),inner sep=5pt]{};
   \node(user) [actor, above=4pt of guiall,minimum width=2cm] {User};
   \draw[vertical con](info) to node[auto]{\textit{interact}}(user);
   \node(export) [actor,left=0.15cm of guiall,text width=0.96cm,yshift=0.4cm,align=left] {pdf tex jpg png};
   \node(print) [actor,left=0.15cm of guiall.south west,text width=0.96cm,yshift=0.2cm] {Printer};
   \draw[->](tree) -- (export);
   \draw[->](tree) -- (print);
   \node(tmodel) [actor,right=1.21cm of info] {Tree Model};
   \node(dmodel) [actor,anchor=north west,yshift=-0.3cm] at (tmodel.south west) {Quantitative Models};
   \draw[->](tmodel) --node[auto]{\textit{extend}}(tmodel|-dmodel.north);
   \node(dclass) [actor,right=0.2cm of tmodel] {Domain Classes};
   \draw[->,]($ (dclass.south)!0.5!(dclass.south west) $) coordinate (a) --
   node[auto]{\textit{derrive}}(dmodel.north -| a);
   \node(model) [above,anchor=south west,inner sep=0.2pt,xshift=10pt] at (tmodel.north west|-gui.south){\textbf{Implementation Model}};
   \node(modelall) [collection,fit=(tmodel)(model)(dclass)(dmodel),inner sep=5pt]{};
   \draw[<->](tmodel) -- node[above,pos=0.74]{\textit{edit}} ++(-1.8cm,0) -- ++(0,-0.45cm) coordinate(b)
   --++(-2cm,0) coordinate(a)--(tree.north-|a);
   \draw[->](b) -- (term.north-|b);
   \draw[<->](dmodel.south-|tmodel.south) -- ++(0,-8.5pt)coordinate(a) --node[below,yshift=0pt]{\textit{edit}}(tree.south|-a)--(tree);
   \node(disk) [shape=cylinder,draw, aspect=0.1,shape border rotate=90,
   right=5pt of modelall] {Disk};
   \draw[<->](info.north)++(1.44cm,0) -- ++(0,21pt)coordinate(a) --node[above]{\textit{load/save}}(disk|-a)--(disk);
   \draw[<-](tmodel.north west) ++(5pt,0) coordinate(b) -- (b|-a);
   \draw[<->](dmodel.east)++(0,5pt) -|node[below,pos=0.36]{\textit{load/save}}(disk);
   \draw[->](dmodel.east) ++(0,-6pt) -- ++(4pt,0) -- ++(0,-0.4cm) --
   ++(-1.6cm,0) coordinate(a) node[below,pos=0.2]{\textit{bottom-up evaluation}} -- (a|-dmodel.south);
\end{tikzpicture}}
\caption{An overview of the ADTool architecture}
\label{fig:arch}
\end{figure}

The application has been written in a modular way with a clear distinction 
between the GUI and the Implementation Model.
An overview of the ADTool architecture is depicted in 
Figure~\ref{fig:arch}. 
The Implementation Model consists of 
the Tree Model (which stores the basic tree structure), 
Domain Classes (defining the implemented attribute domains),
and Quantitative Models (which are derived from 
Domain Classes and contain inserted and computed values). 
The functionality of the tool can easily be extended by defining new 
attributes.
For this purpose, a new Domain Class needs to be created and compiled. 
Domain Classes have been designed to be simple, in order to make it possible 
for a user with 
minimal knowledge of Java to add a new domain. 
Due to the use of Java reflection, 
no recompilation or other modifications of the program are required 
after adding a new Domain Class. 

The ADTool is open source software that runs on all common 
operating systems (Windows, Linux, Mac OS). It is implemented in Java and 
its source code is about $16\,000$ LOC. 
In order to run the program, JDK 6 or later is required. 
Additionally the ADTool depends on the 
following free libraries: abego 
TreeLayout~\cite{abego}, 
implementing an efficient and customizable tree layout algorithm in Java, and InfoNode
Docking Windows~\cite{InfoNode}, 
a pure Java Swing based docking windows 
framework, 
allowing to set up windows in a flexible way and to save and restore their layout.
The ADTool is available for download at
\url{http://satoss.uni.lu/software/adtool/}. It can also be launched 
as an online application that uses the Java Web Start 
technology~\cite{Java_Webstart}.

\section{Conclusion}
The ADTool is free and open source software for the creation and 
quantitative analysis of tree-based security models integrating attack and defense 
components. 
The tool has been extensively tested and has proven to be able to handle 
realistic models of large size. 
Its modular architecture allows for 
an easy extension of the application's capabilities.

The ADTool is currently used in a case study where quantitative threat 
analysis of an electronic voting system is performed. Furthermore, we are 
working on combining the attack--defense tree methodology with Bayesian Networks, to 
make probabilistic reasoning about scenarios involving dependent actions 
possible. Related theoretical findings and newly identified features will be 
implemented in the next versions of the ADTool. 

\bibliographystyle{splncs03}

\end{document}